\documentclass{iopart}
\usepackage{graphicx}
\usepackage{iopams}
\usepackage{setstack}
\newcommand{\be}{\begin{equation}}
\newcommand{\ee}{\end{equation}}
\newcommand{\bk}{{{\bf{k}}}}

\newcommand{\bq}{{\bf{q}}}

\newcommand{\bea}{\begin{eqnarray}}
\newcommand{\eea}{\end{eqnarray}}
\newcommand{\ra}{\rangle}
\newcommand{\la}{\langle}
\newcommand{\upa}{\uparrow}
\newcommand{\dna}{\downarrow}
\newcommand{\bS}{{\bf S}}

\newcommand{\dg}{{\dagger}}
\newcommand{\pdg}{{\phantom\dagger}}

\begin{document}
\title[Controlling local moment formation and interactions in bilayer graphene]{Controlling local moment formation and local moment interactions in bilayer graphene}
\author{Matthew Killi$^1$, Dariush Heidarian$^1$ and Arun Paramekanti$^{1,2}$}

\address{$^1$ Department of Physics, University of Toronto, Toronto, Ontario, \\
Canada M5S 1A7}
\address{$^2$ Canadian Institute for Advanced Research, Toronto, Ontario, \\ Canada M5G 1Z8}
\ead{mkilli@physics.utoronto.ca}  
\date{\today}
\begin{abstract}
We study local moment formation for adatoms on bilayer graphene (BLG) within a mean field theory of the 
Anderson impurity model.  The wavefunctions of the BLG electrons induce strong particle-hole asymmetry and band dependence of the hybridization, which is shown to result in unusual features in the impurity model phase diagram.  We also study the effect of varying the chemical potential, as well as varying an electric field perpendicular to the bilayer;
the latter modifies the density of states of electrons in BLG, and, more significantly, shifts the impurity energy.
We show that this leads to regimes in the impurity phase diagram where local moments can be turned on or off by applying modest external electric fields.
Finally, we show that the RKKY interaction between local moments can be varied by tuning of the chemical potential (as has also been suggested in monolayer graphene) or, more interestingly, by tuning the electric field so that it induces changes in the band structure of BLG.
\end{abstract}
\pacs{75.20.Hr, 75.75-c, 73.20.-r, 75.30.Et}
\submitto{\NJP}
\maketitle

\section{Introduction} 

Single layer graphene hosts a plethora of phenomena that arise from the Dirac-like band dispersion and chirality of its low-energy quasiparticle excitations \cite{Castro-Neto:2009lr, Abergel:2010yq}. It is interesting to explore how these unusual single particle properties impact the physics of adatoms on graphene.  The combination of adatom-graphene hybridization and Hubbard-like interactions on the adatom has been studied in the context of local moment formation \cite{Uchoa:2008lr, Venezuela:2009ys}, Kondo physics \cite{Uchoa:2010kx, Zhu, DellAnna, Wehling:2010fk, Jacob:2010yq}, RKKY interactions \cite{Vozmediano:2005fr, Dugaev:2006zr, Brey:2007lr, Saremi:2007rt, Black-Schaffer:2010uq, Black-Schaffer:2010ys, Sherafati:2010rt}, and adatom positional ordering \cite{Cheianov20091499, Berashevich:2009lr, Shytov:2009vn, Abanin:2010fk}. The study of adatoms on monolayer graphene is also of interest to the nanoscience and quantum computation communities given the possibility to control local moment physics, and adatom-adatom spin and density interactions, by varying the carrier concentration via gating \cite{Wolf16112001}.

\begin{figure}[t]
\hfill
a)\includegraphics[width=5.3cm]{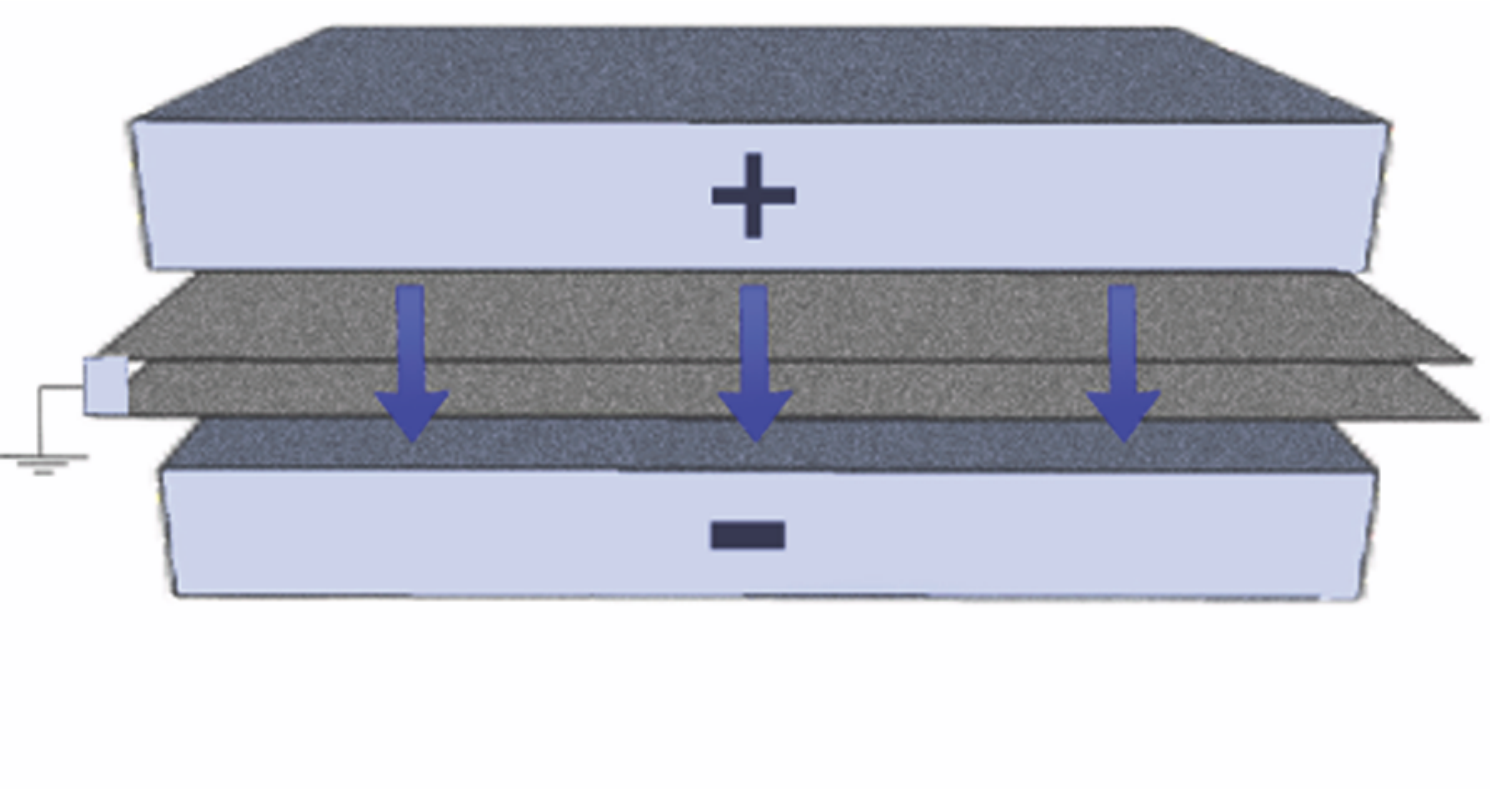}
b)\includegraphics[width=5.3cm]{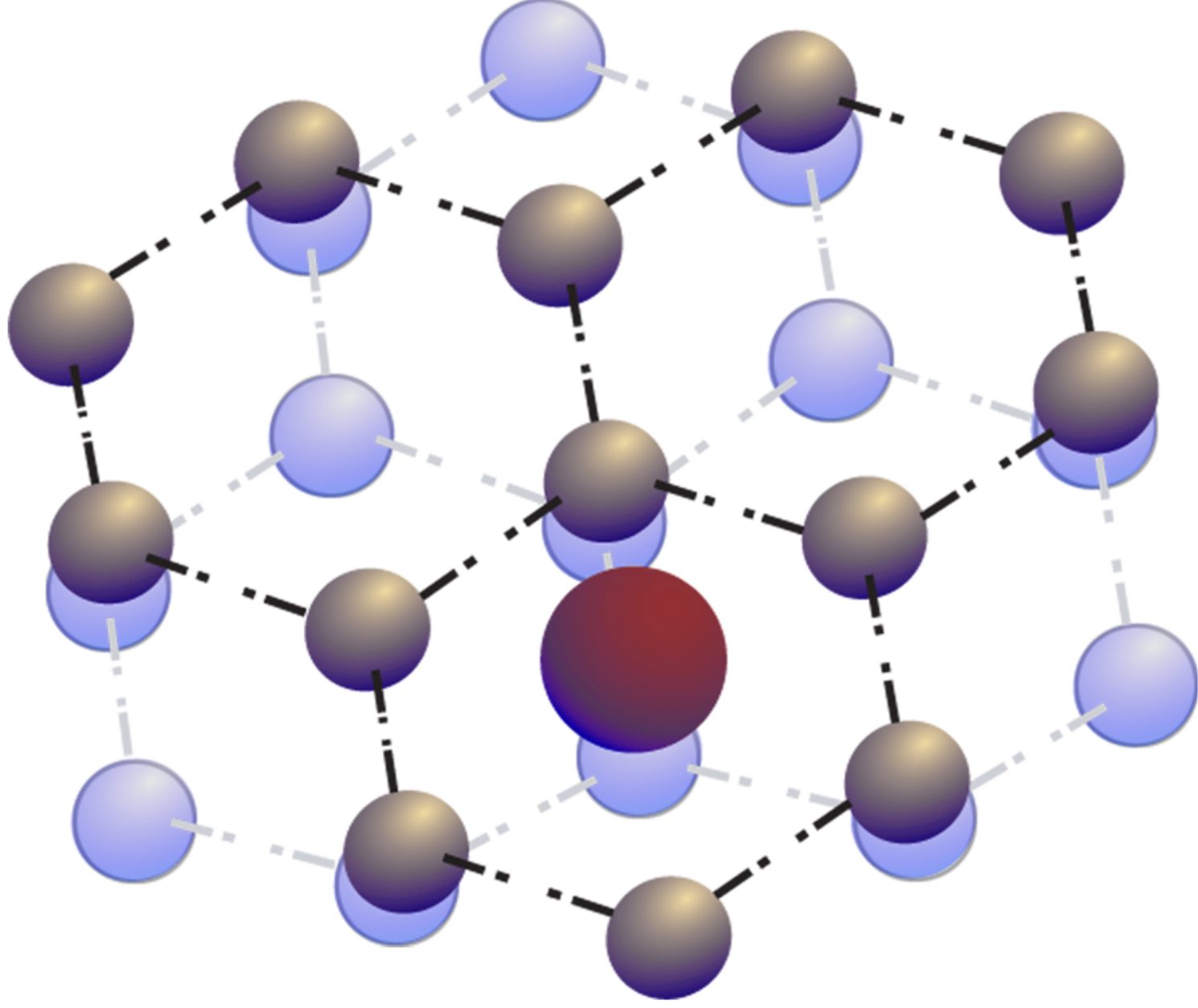} \\

\hfill c)\includegraphics[width=11cm]{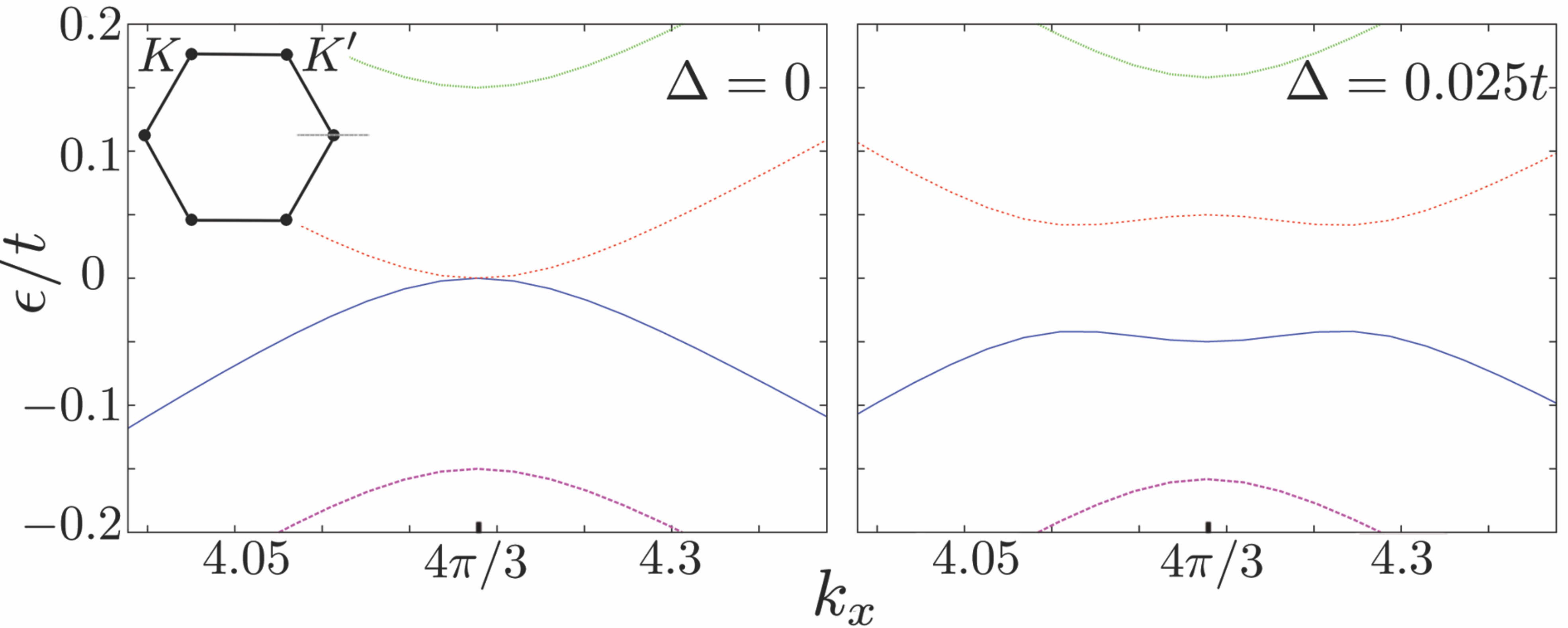} 
\caption{(a) Bilayer graphene in a dual-gate configuration.  (b) Schematic diagram of a plaquette-centered (large, red) adatom impurity on the top layer of bilayer graphene.  (c) Cross-section of the dispersion relation for unbiased (Left) and biased (Right) graphene close along $k_y=0$ through the K-point ($\Delta=0$ and $\Delta=0.025t$, respectively).  Inset:  The two unique K-points and the cross-sectional cut are indicated in the Brillioun zone.}
\label{model}
\end{figure}

In contrast to monolayer graphene, bilayer graphene (BLG), which has Bernal stacking of single
layers, has an extra tuning parameter. Using a dual-gate geometry, shown in figure \ref{model}a, enables one to separately tune the chemical potential and an electric field perpendicular to the layers, which is equivalent to separately tuning the potential on each of the two layers of BLG. While tuning the chemical potential modifies the carrier concentration, applying an electric field normal to the layers generates a gap in the band structure of BLG \cite{McCann:2006qy, Castro:2007lr, Oostinga:2008fk, Zhang:2009fj, Jing:2010, Taychatanapat:2010uq} (Figure \ref{model}c). (We will refer to the potential difference between the two layers, induced by this electric field, as the `bias'.)  Such a tunable gap system enables one to envision device applications and the ability to dynamically control various states in BLG \cite{Castro:2009, Castro:2010fk}   This tunability also allows for the study of interesting fundamental physics --- for instance, it has been shown that engineering the electric field to flip direction (from pointing up to pointing down) as a function of position
leads to localized one-dimensional modes at the kink in the bias \cite{Martin:2008ly, Xavier:2010gf, Nunez:2010ve}.
We have shown in recent work that incorporating interaction effects
converts this `nanowire' into a 2-band Tomonaga-Luttinger liquid whose properties, such as Luttinger
parameters and mode velocities, can be controlled by
the bias strength \cite{Killi:2010ul}.

In this paper, we study adatoms in BLG.  We examine local moment formation on the adatoms, RKKY interaction between such local moments, and how these effects can be controlled by tuning the chemical potential and a applying perpendicular electric field. 

Our work goes beyond Ref.\cite{Ding}, which studied local moment formation for site-centered adatoms
on BLG, in several important
respects.  (i)  We consider adatoms that are positioned at the center of a hexagonal plaquette on one of the layers.  The study of this configuration is motivated by a recent ab initio study of adatoms in monolayer graphene that indicates plaquette centered impurities are generally more energetically favourable than on-site impurities \cite{Chan:2008qy}.  We expect a similar situation to hold in BLG.
(ii) An applied electric field is shown to directly tune the impurity energy. This is because an impurity position will, in general, be located closer to the top layer of BLG.
Accounting for this impurity energy 
shift allows us to identify regions of the phase diagram where local moment formation can be 
turned on and/or off by the application of a perpendicular electric field.  
(iii) For a particular impurity level chosen so that its $renormalized$ (with self-energy corrections) energy level lies in the middle of gap in presence of the bias, we construct phase diagrams at zero, positive and negative bias by sweeping the chemical potential.  The resulting phase diagram exhibits the onset of a Coulomb-blockade phase where any arbitrarily small $U$ results in the formation of local moments.  (iv) As a consequence of the chiral wavefunctions of BLG and the fact that the plaquette centered impurity adatom couples to many sites, the coupling between the impurity and the quasiparticles of BLG has strong momentum and band dependence. This affects many of the details of the phase diagram.  For instance, the self-energy develops a large real part that has nontrivial frequency dependence, and substantially renormalizes the position of the impurity spectral peak in a manner that depends on the chemical potential and the applied bias. We provide detailed a physical explanation for how this affects the resulting phase diagrams, which were not provided in reference \cite{Ding}.  Furthermore, to better illustrate the effect of the wavefunctions and chirality of BLG on the phase diagrams, the BLG system is contrasted with a fictitious system of non-chiral fermions with the same DOS and dispersion relation. (v) We go beyond the issue of local moment
formation to address the tunable RKKY interactions between such local moments on BLG.

We begin, in Section II, by introducing the Anderson impurity model specific to BLG.  Section III summarizes
the Anderson mean field theory formalism.  Armed with this background, in Section IV we construct the impurity model phase diagrams for plaquette-centered adatoms (shown schematically in figure \ref{model}b). To highlight some of the unusual features of these phase diagrams, we contrast it with an impurity model of a fictitious system of electrons that have an identical dispersion but a band-independent coupling to the adatom.  Finally, in Section V, we discuss the RKKY interaction, and its tunability, for local moments on BLG.

\section{Adatom model in bilayer graphene}
\label{formalism}

Consider an adatom on BLG, described by the Anderson impurity model \cite{Anderson:1961lr},
\bea
H_{\rm BLG} &=& \sum_{\bk,s,\sigma} (\epsilon^\pdg_{\bk s} - \mu) c^\dg_{\bk s \sigma} c^\pdg_{\bk s \sigma}, \\
H_{\rm imp} &=& 
\sum_\sigma (\epsilon^\pdg_d - \mu) d^\dg_\sigma d^\pdg_\sigma + U n^\pdg_{d \upa} n^\pdg_{d \dna}, \\
H_{\rm mix} &=& - \sum_{{\bf{r}} \sigma} \chi^\pdg_{\bf{r}} (c^\dg_{{\bf{r}} \sigma} d^\pdg_\sigma + d^\dg_\sigma c^\pdg_{{\bf{r}} \sigma}).
\eea
Here $\epsilon^\pdg_{\bk s}$ is the BLG electron dispersion for
electrons labelled by momentum $\bk$ and band index $s$.
We assume a minimal model for the BLG dispersion that includes a nearest-neighbor hopping amplitude, $t$, to sites on the same layer, and an interlayer hopping amplitude, $t_\perp$, between the two sites that sit one on top of the other.
Henceforth, we set $t=1$ and note that $t \approx 3$ eV and $t_\perp/t \approx 0.15$ in BLG.  In $H_{\rm imp}$, we denote the impurity energy by $\epsilon_d$, while $U$ denotes the electron-electron repulsion on the impurity site. 
BLG electrons at sites ${\bf{r}}$ can hop on or off the adatom impurity with an amplitude $\chi^\pdg_{{\bf{r}}}$.
We assume a common equilibrium chemical potential $\mu$ for the impurity and BLG electrons.  The complete Hamiltonian for unbiased BLG is then given by $H=H_{\rm BLG}+H_{\rm imp}+H_{\rm mix}$.

Electronic structure studies of transition metal adatoms on monolayer graphene suggest that the low-energy configuration of many types of impurities corresponds to the adatom residing at the center of a hexagonal plaquette \cite{Chan:2008qy}. We therefore fix the adatom position to be at the plaquette center on the top layer (labelled $\ell=1$) of BLG, as shown in the schematic diagram on the right in figure \ref{model}. For simplicity, we assume that $\chi^\pdg_{\bf{r}}\!\!=\!\!\chi$ for the set of sites $\{{\bf{r}}_n\}$, which includes the six nearest neighbor plaquette sites in layer-$1$ and the site on layer-$2$ that lies directly below the adatom, and $\chi_{\bf{r}}\!\!=\!\!0$ for all other sites.
This simplifying assumption about the impurity model allows us to focus on unconventional features of local moment formation intrinsic to bilayer graphene. Future density functional studies would be useful in incorporating details of the impurity atomic orbitals.  Turning to the mixing Hamiltonian $H_{\rm mix}$ which allows the impurity electrons to hybridize with the BLG electrons, let us set
\be
\label{eq_Vkn}
V^\pdg_{\bk s} \equiv \chi \sum_{{\bf{r}}=\{\!{\bf{r}}_n\!\}} \phi^\pdg_{\bk s}({\bf{r}}),
\ee
where $\phi^\pdg_{\bk s}({\bf{r}})$ denotes the wave function at site ${\bf{r}}$ for electrons in band-$s$ and 
momentum $\bk$.  We then obtain
\bea
\fl
H=\sum_{\bk,s,\sigma} \left( (\epsilon^\pdg_{\bk s} - \mu) c^\dg_{\bk s \sigma} c^\pdg_{\bk s \sigma}
+V^\pdg_{\bk s} c^\dg_{\bk s \sigma} d^\pdg_\sigma + V^*_{\bk s} d^\dg_\sigma 
c^\pdg_{\bk s \sigma}\right) \nonumber \\
+\sum_{\sigma}(\epsilon^\pdg_d - \mu) d^\dg_\sigma d^\pdg_\sigma + U n^\pdg_{d \upa} n^\pdg_{d \dna} .
\eea
While the impurity model Hamiltonian in BLG looks similar to that in conventional systems or monolayer
graphene, there are two important new ingredients in the impurity physics of BLG with plaquette centered impurities.

\begin{figure}[tb]
\hfill
\includegraphics[width=10.5cm]{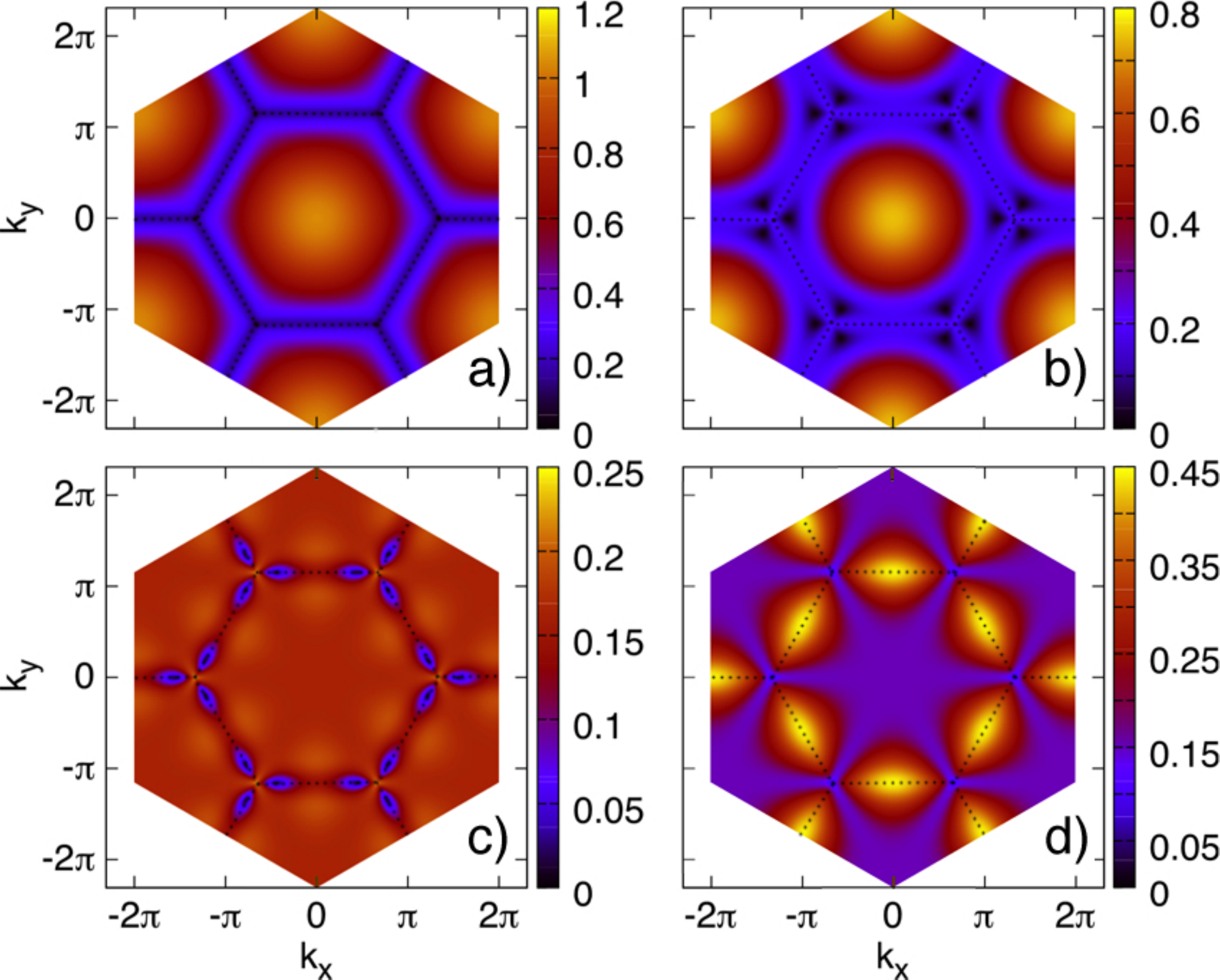}
\caption{ Coupling of the impurity to the four bands (ordered from lowest to highest energy and scaled by system size), 
(a) $|V_{\bk 1}|$, (b) $|V_{\bk 2}|$, (c) $|V_{\bk 3}|$, and (d) $|V_{\bk 4}|$, with impurity
hopping strength $\chi = 0.3t$. Dotted line indicates the Brillouin zone.}
\label{Vkn}
\vskip -0.2in
\end{figure}

First, for BLG (or multilayer graphene), as opposed to monolayer graphene, one can tune the density of states by applying an electric field perpendicular to the layers. Let $\Delta$ denote the potential imbalance between the top and bottom layer induced by the electric field. Assuming that the adatom is at the same height as the top layer, this leads to an extra term in the Anderson Hamiltonian
\be
\label{bias}
H_{\rm bias} = - \frac{\Delta}{2} \sum_{\ell,{\bf{r}}_\ell,\sigma} (-1)^\ell c^\dg_{{\bf{r}}_\ell \sigma} c^\pdg_{{\bf{r}}_\ell \sigma}
 + \frac{\Delta}{2} \sum_\sigma d^\dg_\sigma d^\pdg_\sigma
\ee
where ${\bf{r}}_\ell$ denotes the sites in the top ($\ell=1$) and bottom ($\ell=0$) layers.  In writing this modification to the Hamiltonian, we have assumed that $\chi$ and $t$ remain unchanged in the presence of an electric field.  If intercalation of the impurity occurs, this will reduce the shift in the impurity energy, but will always be nonzero on grounds of the crystal symmetry.
Incorporating the bias in this way thus has three effects: (i) a renormalization of the BLG dispersion; (ii) a modification of the hybridization $V_{\bk s}$ through a  change in the BLG quasiparticle wavefunctions; and (iii) a shift the impurity energy to $\epsilon^\pdg_d +\Delta/2$.  We will refer to the renormalized BLG dispersion and the hybridization as $\epsilon^\pdg_{\bk s}(\Delta)$ and $V_{\bk s}(\Delta)$ respectively.  It is well-known that such a bias in BLG can open a band gap and significantly change the low-energy density of states; what is perhaps not appreciated is that this also effectively tunes the impurity energy in multilayer graphene.  The last term in equation \ref{bias} describing this effect was not present in reference \cite{Ding} and it will be shown to have a remarkable effect on local moment formation in presence of a bias.

A second important difference arises from the tunneling matrix elements, $V_{\bk s}$, for the four bands of the bilayer.  As shown in figure \ref{Vkn}, these matrix elements display strong band- and momentum-dependence, which does not appear for the site-centered impurities discussed in reference \cite{Ding}.  The rich structure of the coupling between the chiral BLG quasiparticles and the impurity site leads to a number of differences in the impurity model phase diagram when compared with conventional non-chiral fermions with a similar density of states, where we simply replace $\phi_{\bk s}({\bf{r}})
\sim \exp(i\bk\cdot{\bf{r}})$ in equation \ref{eq_Vkn}.

\section{Mean field theory}

A mean field treatment of the adatom impurity model is obtained, following Anderson \cite{Anderson:1961lr}, by setting 
\be
U n^\pdg_{d \upa} n^\pdg_{d \dna}  = U  \sum_{\sigma=\pm} 
(\frac{1}{2} \rho^\pdg_d -  \sigma m^\pdg_d) n^\pdg_{d\sigma}
\ee
where $\rho_d = \sum_\sigma \la n^\pdg_{d\sigma} \ra$, and 
$m_d= \frac{1}{2} \sum_\sigma \sigma \la n^\pdg_{d\sigma} \ra$. Let us then define 
\bea
\xi^\pdg_{d\sigma} &\equiv& \epsilon_d - \mu + U (\frac{\rho^\pdg_d}{2} - \sigma m^\pdg_d) \\
\xi^\pdg_{\bk s}(\Delta) &\equiv& \epsilon_{\bk s}(\Delta) - \mu.
\eea
With this mean field approximation, the entire Hamiltonian splits into two single particle impurity
Hamiltonians, one for each spin, with
\bea
H^\sigma_{\rm imp} &=& 
( \xi^\pdg_{d\sigma} + \frac{\Delta}{2}) d^\dg_\sigma d^\pdg_\sigma \\
H^\sigma_{\rm BLG} &=& \sum_{\bk,s} \xi^\pdg_{\bk s}(\Delta)  c^\dg_{\bk s \sigma} c^\pdg_{\bk s \sigma} \\
H^\sigma_{\rm mix} &=& - \sum_{\bk} (V^\pdg_{\bk s}(\Delta) c^\dg_{\bk s \sigma} d^\pdg_\sigma \!+\! 
V^*_{\bk s}(\Delta) d^\dg_\sigma c^\pdg_{\bk s \sigma}).
\eea
These are coupled together by the self-consistency conditions that fix $\xi^\pdg_{d\sigma}$ via $m^\pdg_d$ and $\rho^\pdg_d$. The single particle Green function for the impurity is given by
\be
G_{dd}^\sigma(i \omega_n) = \frac{1}{i\omega_n - (\xi^\pdg_{d\sigma} + \frac{\Delta}{2}) - \Sigma^\pdg_{d} (i\omega_n)},
\ee
where the impurity self-energy is given by
\be
\label{eq_selfenergy}
\Sigma_{d} (i\omega_n) = \sum_{\bk s} \frac{|V_{\bk s}(\Delta)|^2}{i\omega_n-\xi_{\bk s}(\Delta)}.
\ee
We can analytically continue this to the real frequency axis by setting $i\omega_n \!\to\! \omega + i 0^+$ to obtain the
real and imaginary parts of the self-energy $\Sigma_d(\omega)$.  We can then compute at $T=0$
\bea
\rho^\pdg_d \!\!&=&\!\! -\frac{1}{\pi} \int_{-\infty}^0 \!\!d\omega \sum_\sigma {\cal I}m~G^\sigma_{dd} (i\omega_n \!\to\! \omega + i 0^+), \\
m^\pdg_d \!\!&=&\!\! -\frac{1}{2 \pi} \int_{-\infty}^0 \!\!d\omega \sum_\sigma \sigma {\cal I}m~G^\sigma_{dd} (i\omega_n \!\to\! \omega + i 0^+).
\eea
Within this mean field approach, the presence of a local moment on the impurity
is signalled by a self-consistent solution with a 
nonzero $m^\pdg_d$.

Alternatively, it is possible to self-consistently solve the mean field Hamiltonian using exact diagonalization for small system sizes.  All of the phase diagrams in the next section were checked for consistency using this method.

\section{Local moment formation}

Using the above mean field theory enables us to study local moment formation on an impurity atom residing on BLG. 
Since the BLG band structure can be tuned by the electric field, we choose to define $\Gamma_0\! \equiv \! \pi \chi^2/t$ 
as a rough scale for the impurity level broadening in the absence of interactions.  Thus, $\Gamma_0$ remains fixed
for a given $\chi$ even as the electric field and chemical potential are varied.  In this section, we begin by discussing the case when $\Delta=0$ (i.e.\ without an applied electric field perpendicular to the layers).  Phase diagrams are constructed by varying $\epsilon_d$ and $U$ for fixed $\chi=0.3 t$ (which implies $\chi \sim 1$ eV in conventional units) with various choices of the chemical potential.  Next, we consider how varying $\Delta$ can be used to tune the phase diagrams.  After which, we discuss an alternative phase diagram for an impurity with a fixed bare energy level (although the actual energy level will be modified in the presence of a bias) with various choices of $\Delta$.  To construct these phase diagrams, $\mu$ and $U$ are varied, while $\epsilon_{d}$ and $\chi$ ($=0.3t$) are kept fixed.

We have checked that varying $\chi$ modestly makes no qualitative changes to various features in the phase diagram, although it does shift the phase boundaries as expected. We ascribe the complexities of the impurity model phase diagram in BLG to the effective momentum- and band-dependent mixing $V_{\bk s}$. As we discuss below, the strong variation of this coupling between different bands results in particle-hole asymmetry of the impurity model phase diagram via the impurity self-energy. This is despite the fact that in the simplest tight-binding parameterization, which we have considered, the BLG band dispersion itself is particle-hole symmetric for $\mu=0$.  

\subsection{Phase diagram in the unbiased case: $\Delta=0$}
\begin{figure}[t]
\hfill
\includegraphics[width=10.5cm]{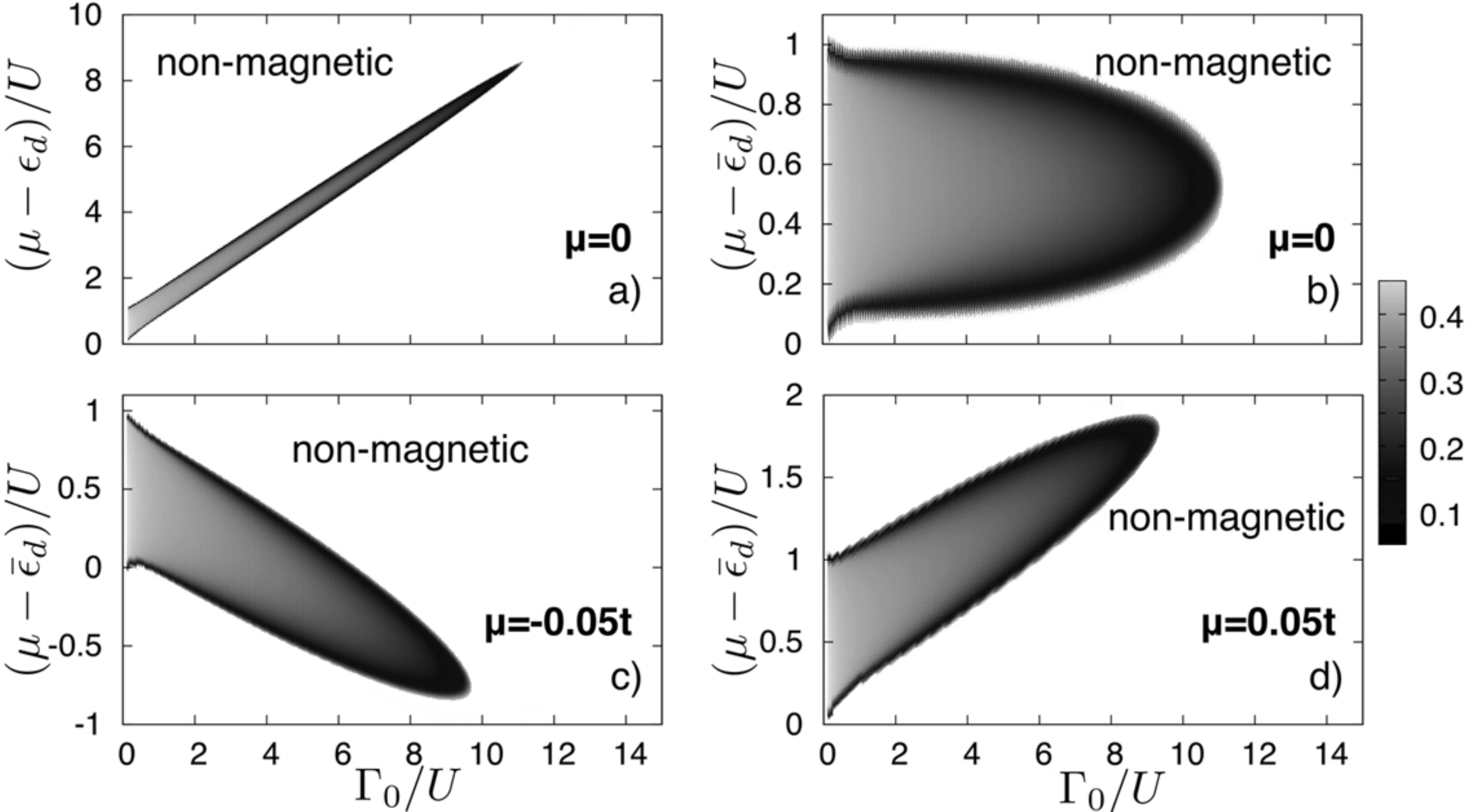}
\caption{Phase diagram of local moment formation on plaquette centered impurities in terms of $\epsilon_{d}$ for 
(a) $\mu=0$, and in terms of $\overline{\epsilon}_{d}=\epsilon_{d} +\Sigma'_{d}(0)$ for (b) $\mu=0$, (c) $\mu=-0.05t$, 
and (d) $\mu=0.05t$ d).  $\chi=0.3t$ in all figures. Grey-scale measures the local moment $m_d$.}
\label{plaquette}
\end{figure}

The $T\!=\!0$ mean field phase diagram for a plaquette-centered impurity embedded in `intrinsic' ($\mu=0$) bilayer graphene with $\Delta=0$ is shown in figure \ref{plaquette}(a). The phase diagram shares some qualitative features with that of local moment formation in a typical host metal.  Namely, there exists a critical ratio of $\Gamma_0/U$ before the onset of mean field magnetization and a clear Coulomb staircase in the small $\Gamma_0/U$ limit.  Despite these similarities, there are two unusual aspects to this phase diagram.  We next start by highlighting these novel features and then clarify their physical origin.

(i) As seen from figure \ref{plaquette}(a), there is an extreme skewing of the magnetic regime from being centered at $(\mu-\epsilon_{d})/U \! \sim \! 0.5$ for small $\Gamma_0/U$ to being centered around large positive values of $(\mu-\epsilon_{d})/U$ with increasing $\Gamma_0/U$. This strong particle-hole asymmetry arises from the fact that the impurity couples asymmetrically to the two layers of BLG, leading to a significant real part of the impurity self-energy $\Sigma_{d}'(\omega)$. The effect of which is to strongly renormalize $\epsilon_d$, which causes the observed skewing.   In order to eliminate this large skewing in later plots, we split the real part of the impurity self-energy as
\be
\Sigma_{d}'(\omega) =  \Sigma_{d}'(0) + (\Sigma_{d}'(\omega) - \Sigma_{d}'(0)) 
\ee
and absorb $\Sigma_{d}'(0)$ into the impurity energy, defining a renormalized impurity energy
$\bar{\epsilon}_{d}=\epsilon_{d}+\Sigma'(0)$.
The resulting renormalized self-energy
$\tilde{\Sigma}_{d}'(\omega) = (\Sigma_{d}'(\omega) - \Sigma_{d}'(0))$ then vanishes at $\omega=0$, 
and remains small but nonzero away from $\omega=0$.  Plotting the impurity model phase diagram in terms of the renormalized impurity energy $\bar{\epsilon}_{d}$, to a large degree but {\it not completely},  removes the strong particle-hole asymmetry for $\mu=0$; this can be seen in figure \ref{plaquette}(b). Of course, strong particle-hole asymmetry continues to exist away from $\mu=0$ even after accounting for the impurity energy renormalization, as shown in figure \ref{plaquette}(c),(d); this can be ascribed to the particle-hole asymmetry of the BLG dispersion at nonzero $\mu$.

(ii) As seen from figure \ref{plaquette}(a), there is a dramatic elongation of the magnetic region to large values of $\Gamma_0/U \sim 10$, which one can partially attribute the small density of states at $\mu=0$.  However, the phase diagram is also influenced by the {\it wavefunctions} of the BLG quasiparticles.  A close inspection of the phase boundaries reveals that they are not symmetric about $\mu=0$ even after accounting for the self-energy correction discussed above.  We understand that this residual particle-hole symmetry breaking arises from the asymmetric broadening of impurity level caused by the disparate effective hybridizations with the different bands. This effect is also seen in the phase diagrams for systems by comparing the $\mu=0.05t$ and $\mu=-0.05t$ phase diagrams.  While one might na\"ively expect that the symmetry between the valence and conduction dispersions would lead to symmetric phase diagrams for positive and negative chemical potential, subtle di
 fferences between the two regions again reflect the influence of the wavefunctions of the electrons that hybridize with the impurity level.

It is, in fact, extremely instructive to compare the complete impurity phase diagram of bilayer graphene with a fictitious system of electrons obtained by setting $\phi_{\bk s}({\bf{r}}) = \exp(i\bk\cdot{\bf{r}})/\sqrt{N_s}$ in equation \ref{eq_Vkn}, where $N_s$ is the total number of sites in the bilayer.  These fictitious electrons are chosen to have the same dispersion as the BLG quasiparticles, but their coupling to the impurity does not account for the chirality or the band dependence of the quasiparticle wavefunctions. We find that some of the unusual features of the BLG impurity phase diagram, discussed above, are eliminated upon making this change.

\begin{figure}[t]
\centering
\includegraphics[width=7.5cm]{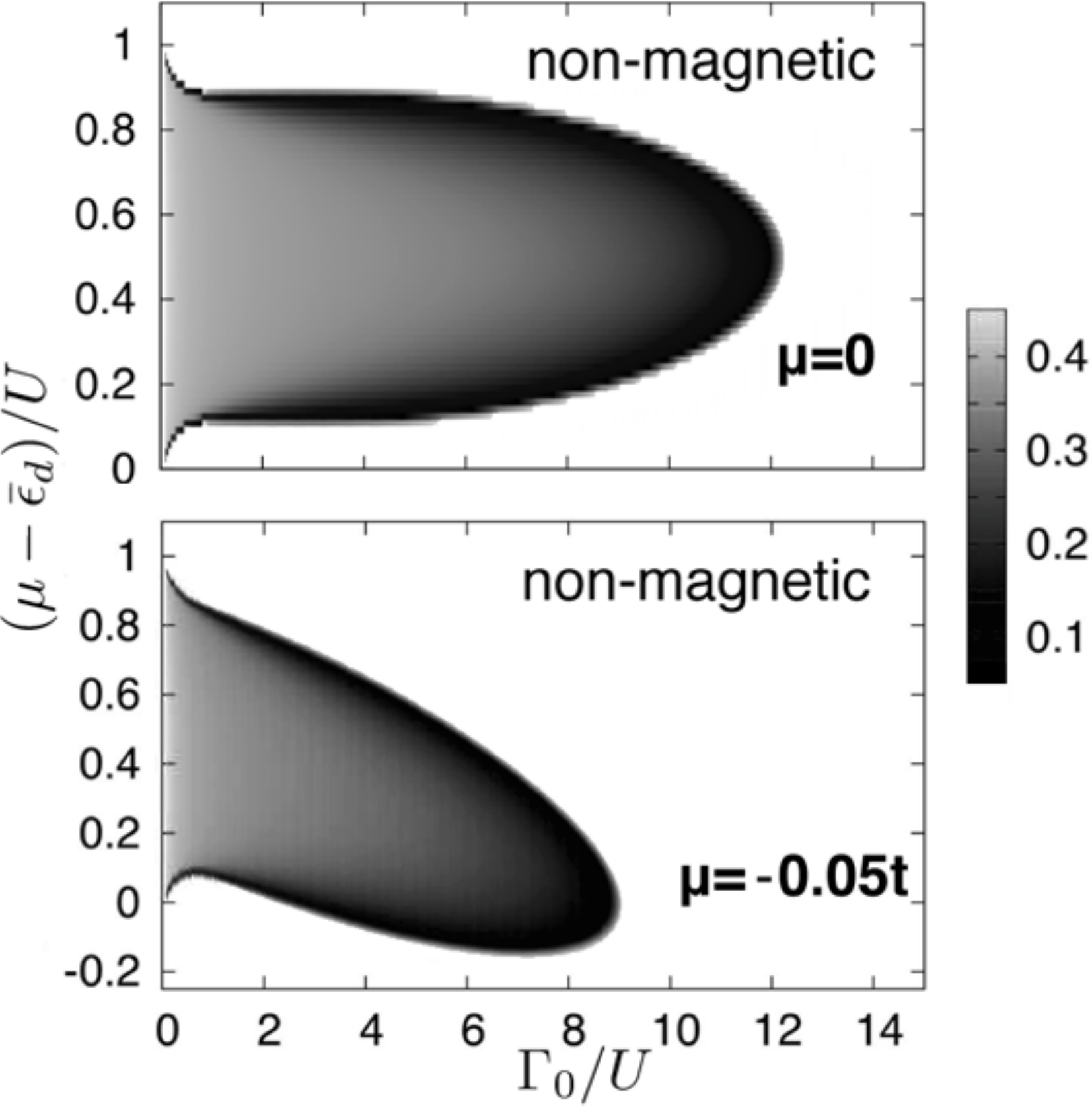}
\caption{The phase diagram of local moment formation for fictitious fermions with the same dispersion as bilayer graphene for (above) $\mu=0$, and (below) $\mu=-0.05t$, plotted in terms of $\bar{\epsilon_d}$. (Note, $\bar{\epsilon}_d=\epsilon_{d}$ when $\mu=0$.)  The phase diagram for $\mu=0.05t$ is related to that of $\mu=-0.05$ by a reflection about $\mu-\bar{\epsilon}_{d}/U=0.5$.  }
\label{fiction}
\end{figure}

Most noticeably, the phase diagram of the fictitious fermions is not skewed when $\mu=0$ even when plotted in terms of the unrenormalized impurity energy, indicating that $\bar{\epsilon}_d=\epsilon_d$, so that
$\Sigma_d(0)=0$.  This stems from a symmetry $\Sigma_d(-\omega)=-\Sigma^*_d(\omega)$ in the expression for the self-energy in equation \ref{eq_selfenergy} upon assuming a band-independent $V_{\bk s}$.  Moreover, it also follows that the
phase diagram for the fictitious fermions is {\it exactly} particle-hole symmetric, in contrast to the case of BLG.

Similar arguments also exactly relate the non-chiral phase diagrams of systems with corresponding chemical potential 
$\mu$ and $-\mu$ by noting that the self-energy at finite chemical potential can be obtained by $\Sigma_{d}(\omega+\mu)$ 
of the self-energy at $\mu=0$.  Consequently, the phase diagram of the $-\mu$ system is obtained by reflecting the phase diagram of the $\mu$ system.  This is again in contrast to the phase diagram of BLG where there is no such relation between systems with positive and negative $\mu$.  In BLG, particle-hole excitations in a system with positive chemical potential favour different bands than those of a system with negative chemical potential.  Since each band has a unique effective coupling to the impurity in BLG, the hybridization of the impurity states will depend on the sign of the chemical potential and so the phase diagrams will be different.  Finally, the other major distinction between the finite chemical potential phase diagrams of the two systems is that, once again, the BLG phase diagram is more strongly skewed, even when plotted in terms of $\bar{\epsilon}_{d}$.  This confirms that the band- and momentum-dependence of the hybridization to the
BLG quasiparticles is responsible for sizeable shift in the impurity energy via a sizeable real self-energy.

\subsection{Phase diagram in the biased case: $\Delta \neq 0$}

We now turn our attention towards a BLG system in a dual-gate configuration.  This setup allows one to continuously tune the layer bias and the average chemical potential independently by applying an external electric field perpendicular to the layers.  In the presence of a symmetric interlayer bias, the chemical potential remains fixed while a band gap opens in the bulk electronic spectrum of BLG.  In the context of local moment formation, this modification to the density of states is expected to substantially change the extent to which an impurity state hybridizes with the BLG electrons.  In addition to this, the impurity energy levels also shift up or down depending on the potential of the layer in which it resides.  This remarkable ability to alter the energy of an impurity level with respect to the chemical potential through the application of an external electric field is unique to multilayer systems, and has no analog in monolayer graphene.

In the first part of this section, we explore how biasing the layers affects local moment formation by reconstucting phase diagrams similar to those above, but for gated systems with different layer bias and fixed $\mu=0$.  Doing so allows us to identify regions of impurity parameters where local moment formation can be turned on and/or off by the electric field.  In the subsequent part of this section, we consider the ability to tune both the chemical potential and bias by constructing alternative phase diagrams where $\mu$ and $U$ are varied and it is the bare impurity energy which is fixed.  This is again done for a selection of values for the bias. 

\subsubsection{Impurity energy variation}

\begin{figure}[t]
\centering
\includegraphics[height=8cm]{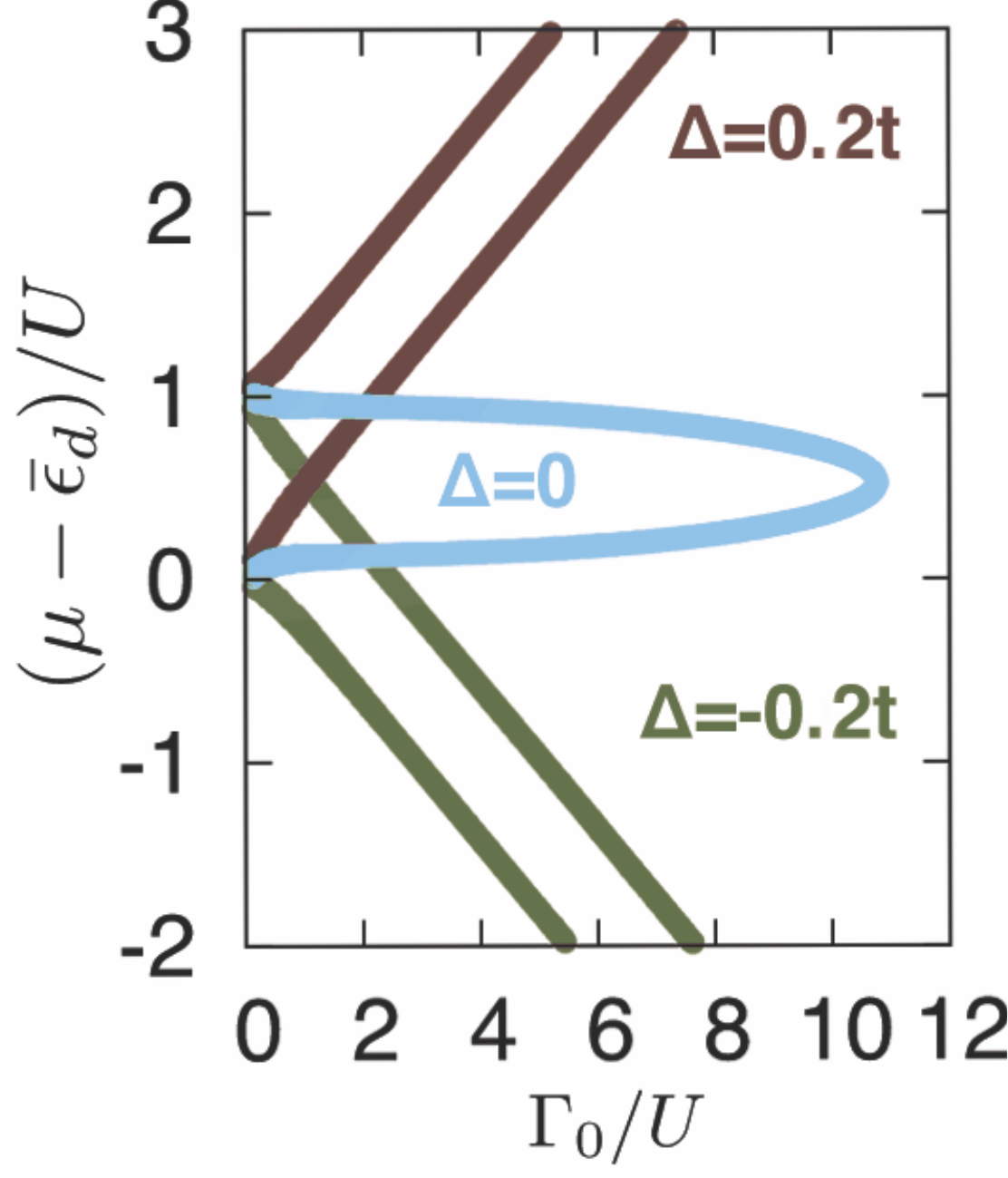} 
\caption{Phase diagram of local moment formation in plaquette centered impurities with $\Delta=0.0t$, $0.2t$, $-0.2t$ as a function of $\overline{\epsilon}_{d}=\epsilon_{d} +\Sigma'_{d} \left(\omega=0, \Delta=0\right)$. In both figures $\chi=0.3t$ and  $\mu=0$.}
\label{phaseV}
\end{figure}

Figure \ref{phaseV} is the phase diagram of the impurity model for experimentally accessible values of $\Delta$ plotted in terms of the redefined impurity energy $\bar{\epsilon}_{d}$ introduced above ($\chi=0.3t$ and $\mu=0$).  The bias has two effects on the impurity model: (i) it opens a band gap $\sim \Delta$ in the BLG dispersion, and (ii) it shifts the impurity energy by $\Delta/2$. Let us discuss, in turn, the impact of these two effects on the phase diagram.

{\bf (i)} First consider restricting the effect of turning on a bias to opening a gap in the BLG spectrum so that the impurity energy level remains unaltered.  Then, the dominant effect of a large $\Delta$ is the elongation of the phase boundary to large $\Gamma_0/U$, regardless of the parity of $\Delta$.  This occurs because the bias induces a large band gap and, when the impurity spectral peak lies in this gap, the coupling between the impurity and the extended states becomes negligible because the density of states vanishes.  (We have to be careful that the {\it renormalized} impurity energy, taking self-energy corrections into account, should lie in the gap; this renormalization is small if the band gap is large compared to $\Gamma_0$.)  Hence, the impurity spectral functions become simple delta functions and if we vary $\epsilon_d$ for fixed $\Gamma_0/U$ the local moment phase boundary resembles that of a simple Coulomb staircase in the atomic limit.

{\bf (ii)} The effect of shifting the energy of the impurity level is similar to the effect of the real part of the self-energy in the phase diagram; it dramatically skews the local moment phase about $\epsilon_{d}/U=0.5$.  The direction of the skewing depends on the parity of the bias, as this determines the direction of the impurity energy shift.

If the impurity energy shift and opening of a band gap are taken together, both skewing and elongation of the local moment phase boundary occur.  As the electric field is increased from zero to large field strengths, the local moment phase continuously elongates and `peels' away from the zero bias boundary.  Although slight, it is important to note that the phase diagrams with opposite bias parity are not symmetric but have slight differences that arise from the breaking of layer symmetry by the impurity.  One of the key new results is the identification of regions in the impurity parameter space where local moments can be turned either on and/or off by adjusting the electric field.  The region where local moments survive both in the presence and absence of the electric field are simply where the phases overlap.

\subsubsection{Chemical potential variation}

Now we explore the possibility of tuning the chemical potential of the system to control local moment formation both in the unbiased and biased cases.  To do this, we construct phase diagrams for a given $\epsilon_d$ and $\Delta$, and we now vary $\mu$ and $U$.  We do this for $\Delta=0,\pm 0.2 t$, for a choice of the bare impurity energy such that the
noninteracting impurity spectral peak appears in the midgap when $\Delta= -0.2 t$, which we do by choosing 
$\epsilon_{d} + \Delta/2 = - \Sigma(\omega=0,\Delta)$.

\begin{figure}[t]
\center
\includegraphics[height=5.cm]{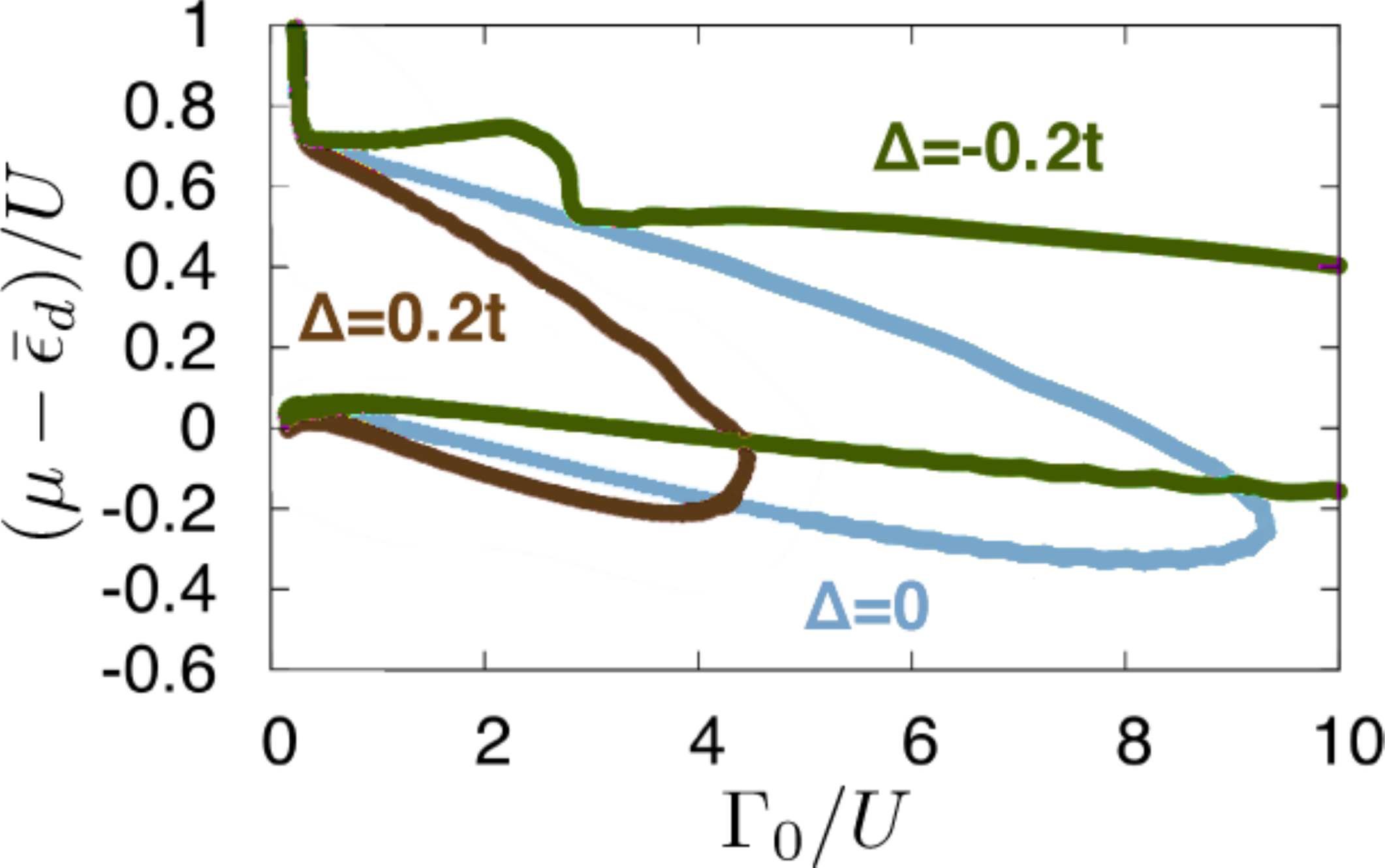}
\caption{Local moment phase diagram for biased bilayer graphene.  The impurity level at $\Delta=-0.2t$ bias was chosen so that its spectral peak lies in the middle of the gap.}
\label{altphase}
\end{figure}

In figure \ref{altphase}, the phase diagram is plotted in terms of a redefined impurity energy $\bar{\epsilon}_{d}
= \epsilon_{d} + \Sigma(\omega=0,\Delta=0)$.  It is important to emphasize that the location of the spectral peaks mostly do not correspond to 
$\bar{\epsilon}_{d}$. The real part of the self-energy has significant frequency dependence that shifts the location of the spectral peak, whose effect must also be accounted for in order to fully understand the phase diagrams.

When the system is unbiased (i.e.\ $\Delta=0$) the impurity energy level lies within the conduction bands (see reference \cite{Castro-Neto:2009lr} or reference \cite{Castro:2007lr} for details on the band structure).  In this case, the phase diagram is qualitatively similar to that of a single site impurity (see reference \cite{Ding}).  When a positive bias is in 
place, $\Delta=0.2 t$,
a band gap opens and the impurity energy shifts deeper into the conduction band.  Consequently, the phase boundaries for local moment formation are significantly reduced because of the enhanced broadening due to the 
increase in the density of states at higher energy in the conduction bands.

The more  interesting case is when $\Delta=-0.2t$ and the impurity spectral peak shifts down into the middle of the gap.  Then, if $U$ is small enough so that the doubly occupied state also lies at sub-gap energies, both the singly and doubly occupied states can no longer hybridize with the BLG states and the impurity spectral function reduces to delta functions.   Hence, we again recover local moment formation very similar to the atomic limit, but now in the large $\Gamma_{0}/U$ limit.  However, in this limit the upper and lower phase boundaries of the Coulomb-staircase are not separated by $(\mu-\bar{\epsilon}_{d})/U=1$ because of level repulsion.  The doubly occupied state shifts down in energy due to $\Sigma'_{d}(\omega)$. 

Thus, this phase diagram is unusual in the sense that it has two regimes resembling the atomic limit at large and small $\Gamma_{0}/U$.  Separating these regimes is the part of the phase diagram where the doubly occupied state's energy lies beyond the band edge and hybridizes with the conduction states.  This occurs at about $\Gamma_{0}/U \sim 2.5$ when $U\sim 0.1t$, precisely where the unusual `hump'-like feature is seen in the upper phase boundary.  The cause of the feature can again be attributed to level repulsion, as it becomes very strong for states close to the the gap edge and $\Sigma'(\omega)$ exhibits a large peak.

\section{RKKY interaction between local moments}
\label{rkky}

In this Section, we explore the RKKY coupling between local moments \cite{Ruderman:1954kx, Kasuya:1956yq, Yosida:1957fj} and study how it can be tuned by varying the band gap and chemical potential using a dual-gate configuration. We have seen in the previous section that such variations will, in general, modify the local moment. Here, we focus on changes to the RKKY coupling induced purely by changes in the bulk band structure and filling.

We consider two classical local moments that couple to the set of sites $\{{\bf{r}}\}$ and $\{{\bf{r}}'\}$, respectively,
\be \label{moments}
H'=\sum_{\{{\bf{r}}\}}J^{(1)}_{{\bf{r}}} \bS_{1}\cdot {\bf s}_{{\bf{r}}}+\sum_{\{{\bf{r}}'\}}J^{(2)}_{{\bf{r}}'} \bS_{2}\cdot {\bf s}_{{\bf{r}}'}, 
\ee
where $J^{(a)}_{{\bf{r}}}$ is the strength of the exchange coupling of an electron's spin, ${\bf s}_{{\bf{r}}}$, at site ${\bf{r}}$ with the magnetic impurity $\bS_a$.  Upon integrating out the itinerant electrons and retaining only those terms that are second order in $J^{(a)}_{\bf r}$, one obtains a reduced Hamiltonian for the local moments,
\be
H_{eff}=J_{RKKY}\bS_1\cdot \bS_2.
\ee
The coupling $J_{RKKY}$ is given by
\bea
\label{J}
\fl
J_{RKKY}=\!
\frac{1}{2N} \! \sum_{\substack{ \bq \bk i j n m }}   M_{i j}(\bq) \,
 \phi^{*n}_{\bk}(i) \phi^{n}_{\bk}(j) \phi^{*m}_{\bk + \bq}(j) \phi^{m}_{\bk + \bq}(i)\, e^{i\bq\cdot({\bf{r}}_{1}-{\bf{r}}_{2})}  \nonumber \\
\times  \frac{n_{F}(\xi^{m}_{\bk+\bq})-n_{F}(\xi^{n}_{\bk})}{\xi^{m}_{\bk+\bq}-\xi^{n}_{\bk}},
\eea
where $m/n$ are band indices, $i/j$ are the combined sublattice and layer label, $n_{F}$ is the Fermi distribution, and $M_{ij}(\bq)$ is a matrix describing the Fourier transform between different sites weighted by $J^{(1)}_{\bf r} J^{(2)}_{\bf r'}$.  The explicit form of $M_{ij}$ for the case of interest is provided below.

For monolayer graphene, it has been shown that a perturbative treatment in the continuum low-energy theory \cite{Saremi:2007rt} produces approximate results that match closely with exact diagonalization \cite{Black-Schaffer:2010uq} and lattice Green's functions methods \cite{Sherafati:2010rt}, as long as an appropriate high-energy cutoff scheme is applied.  In the above perturbative treatment, the entire band structure is used in the calculation so as to avoid any cutoff dependence and the RKKY coupling is accurately reproduced for monolayer graphene.  We therefore expect this perturbative calculation to also be a reasonable approach to study the RKKY coupling in BLG in the dual-gate configuration.

\begin{figure}
\center
\includegraphics[height=5.cm]{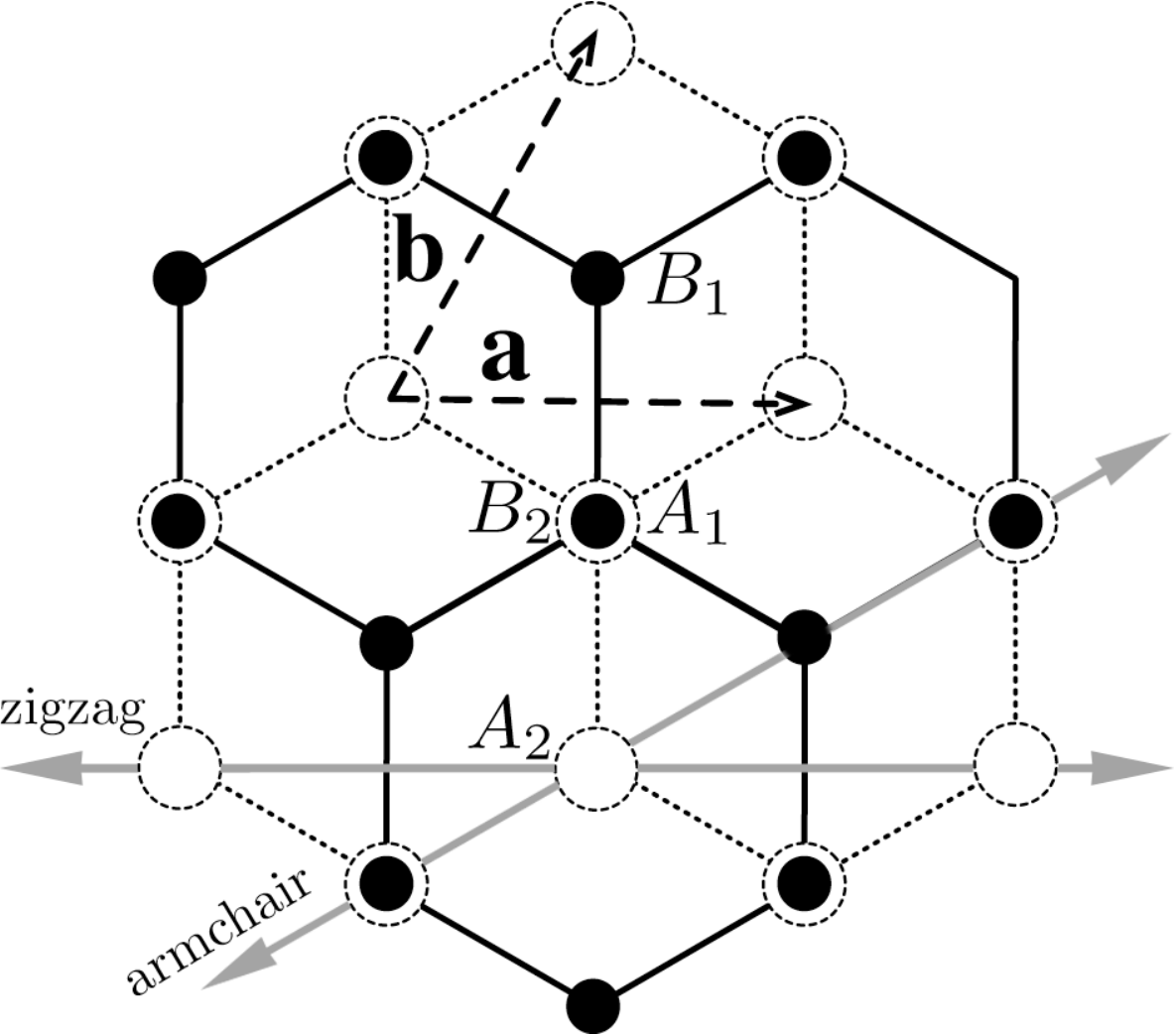}
\caption{Crystal structure of bilayer graphene and site labelling convention.  The primitive lattice vectors are ${\bf a}$ and ${\bf b}$ and the armchair and zigzag directions are indicated by the arrows.  The local moments considered here are plaquette centered and reside on the same layer above an $A_2$ atom.}
\label{crystal}
\end{figure}

We analyzed various moment configurations such as single site (AA, BB, AB) and plaquette coupled moments both along the zigzag and armchair directions (see figure \label{crystal} for labelling conventions).  The effects of varying the chemical potential and layer bias were seen to be qualitatively similar for each case, so we have chosen to present only the results for plaquette centered moments that lie along the zigzag direction.  The impurity atom is taken to lie above an $A_2$ in the center of a hexagonal plaquette in layer 1.   For simplicity, we assume the coupling to each of the seven sites is equal so that $J_{\bf r}^{(1)}=J_{\bf r}^{(2)}\equiv J$, although we have checked that the results are qualitatively unaffected if there is an unequal coupling to the site below the impurity on the other layer ($A_2$).  In this case, the components of $M_{ij}$ are
\bea
M_{A_{1}A_{1}}\!=J^2 \Big( 4+2\left( \cos(q_{a}) + \cos(q_{b}) +\cos(q_{a}-q_{b})\right) \Big)  \nonumber \\ 
M_{A_{1}B_{1}}\!=J^2\Big( 2 + 2\left( \cos(q_{a}) + \cos(q_{b}) +e^{i(q_{a}-q_{b})} + e^{-i(2q_{a}-q_{b})} \right) \nonumber \\
M_{A_{1}A_{2}}=J^2\Big(1+e^{iq_{b}}+e^{i(q_{a}-q_{b})}\Big)   \nonumber \\
M_{B_{1}A_{2}}\!=J^2\Big(e^{iq_{b}}+e^{i(q_{a}-q_{b})}+  e^{i(q_{a}-2q_{b})}\Big)   \nonumber \\
M_{A_{2}A_{2}}\!=J^2, \phantom{\Big( 2}  M_{i j}=M^{*}_{j i},
\end{eqnarray}
where $q_a={\bf q} \cdot {\bf a}$, $q_b={\bf q} \cdot {\bf b}$ and ${\bf a}={\bf \hat{x}}$ and ${\bf b}={\bf \hat{x}}/2 + \sqrt{3}{\bf \hat{y}}$/2.

\begin{figure}[t]
\center
\includegraphics[height=5.cm]{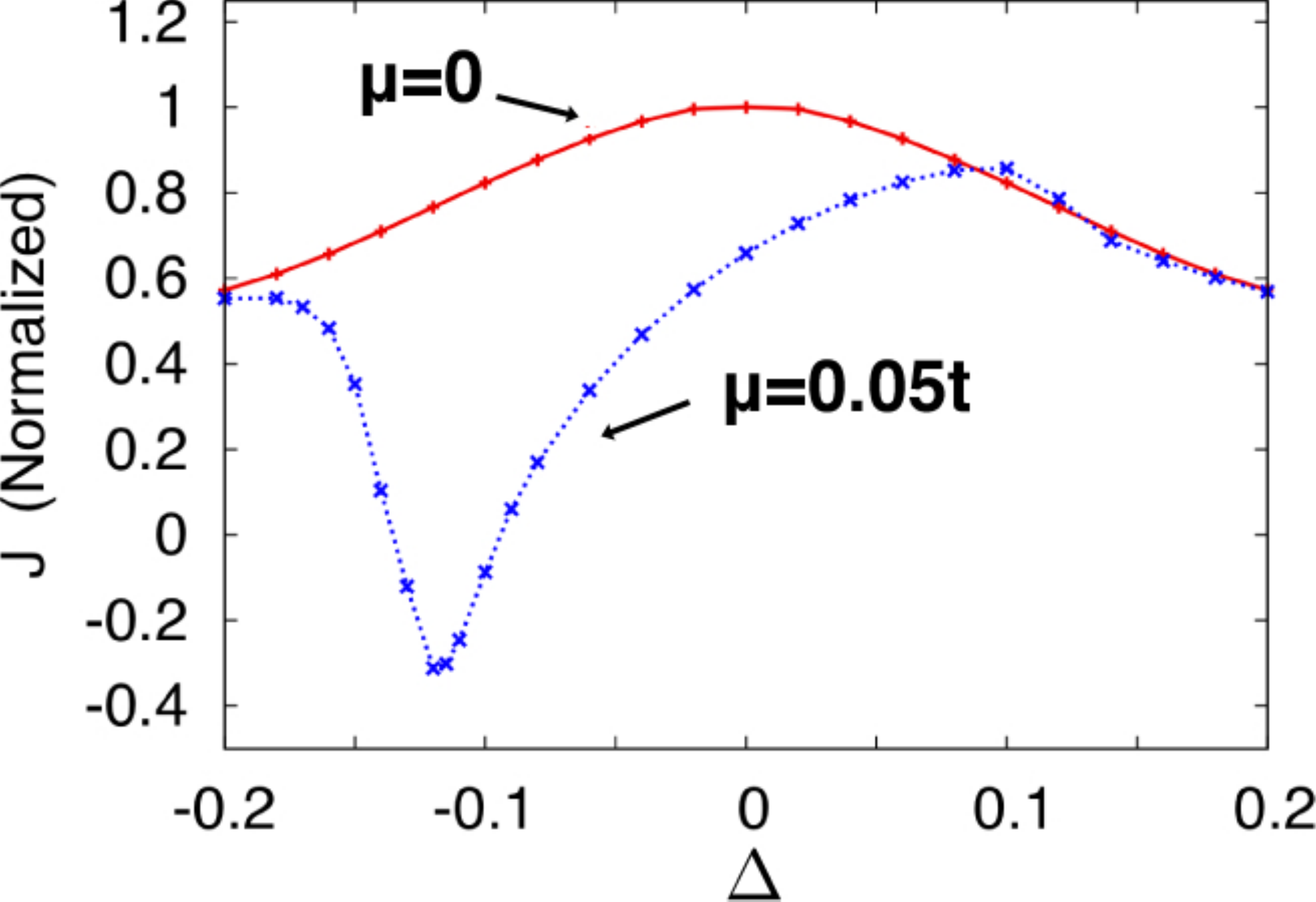}
\caption{Normalized RKKY coupling strength between to classical plaquette centered moments at a distance of $10$ lattice spacings along the zigzag direction.  Both moments are located on the same layer and the chemical potential is chosen to be  $\mu=0$ and $\mu=0.05t$ at a temperature $T=0.002t$ ($\sim70$ K).}
\label{JVvariation}
\end{figure}

To demonstrate the ability to tune the RKKY interaction using the dual-gate configuration, the $J_{RKKY}$ coupling, normalized to its value at $\mu=0$ and $\Delta=0$, is plotted in figure \ref{JVvariation} as a function of the interlayer bias $\Delta$ for two moments separated by 10 lattice spacings.  This is done for $\mu=0$ and $\mu=0.05t$.  For experimental considerations, one must keep in mind that the RKKY coupling is quite small in BLG.  As an example, a bare exchange term equal to $J^{(1)}=J^{(2)}=0.2t$ produces an effective coupling $J_{RKKY}=1.3\times10^{-4}t$ ($\sim 4.4$ K) at 4 lattice spacings, and just $J_{RKKY}=7.5\times10^{-6}t$ ($\sim 0.3$ K) at $10$ lattice spacings.  However, similar to monolayer graphene, electron interactions are expected to make the coupling strength more long ranged \cite{Black-Schaffer:2010ys}.  At shorter distances, the RKKY interaction is enhanced, but the tunability is reduced. 

Before describing the tunable features of the RKKY coupling, it is important to first understand that the wavefunctions of a given band are sensitive to the parity of the bias between the layers, even though the dispersion is not.  Their dependancy on the parity can significantly influence how $J_{RKKY}$ changes with bias, as explained below.  When a positive bias is present, states in the upper two bands are more heavily weighted to layer 1 sites, while states in the lower band are more heavily weighted to the layer 2 sites.  This weighting is reversed when the parity of the bias is negative.  In contrast, when there is no bias the weighting of the wavefunction is the same for each layer.

With this background, it is possible to explain the symmetry/asymmetry between the two curves.  When $\mu=0$, the chemical potential lies between the valence and conduction bands, and so particle-hole excitations can only occur between them.  This corresponds to one of the states being localized to layer 1 and the other localized to layer 2.  It follows that the coupling strength $J_{RKKY}$ is parity invariant and so it is symmetric for positive and negative biases.  

In contrast, when $\mu \neq 0$, the coupling is sensitive to the parity of the bias.  If $\mu=0.05t$, $\Delta>0$ and $\mu$ is less than the band gap, the chemical potential lies in the third band where the states tend to localize to layer 1, the layer in which the moments reside.  The finite chemical potential causes some of the particle-hole excitations between the lower and upper two bands to be suppressed by Pauli-blocking, and also introduces low-energy excitations between the two upper bands where the wavefunctions are weighted to layer 1.  If however, $\mu=0.05t$, $\Delta<0$ and $\mu$ is less than the band gap, the chemical potential lies in the third band, but now these states tend to localize to layer 2.  Although the energetics of the scattering processes remain the same, the matrix elements do not.  The excitations between the upper two bands now have matrix elements whose weighting on layer 1 is much less.  Thus, the coupling is dependent on the $relative$ parity o
 f $\mu$ to $\Delta$.  Hence, if we consider a system with $\mu=-0.05t$, the $J_{RKKY}$ curve will be reflected about $\Delta=0$.

In addition to the effects described above, the density of states about the chemical potential tends to increase for small biases, as the band edge flattens and is pushed closer to the chemical potential.  At large bias strengths, the dispersion close to the band-edge resembles that of a `mexican-hat', leading to further complexity in the density of states.  Furthermore, at finite chemical potential, the Fermi points extend out to form a Fermi-surface symmetric about the $K$-points.  The combination of all these effects lead to the non-trivial changes seen the RKKY coupling in figure \ref{JVvariation}.

Interestingly, at a distance of 10 lattice spaces, the coupling remains antiferromagnetic when $\mu=0.05t$ and $\Delta >0$ and tends to increase with bias strength.  However, when $\Delta$ becomes increasingly negative, the antiferromagnetic coupling strength is reduced to zero then switches to a ferromagnetic coupling about $\Delta \sim -0.12$.  For the case of $\mu=0.05t$ considered here, the ability to fully turn off the coupling and switch the sign of $J_{RKKY}$ with an applied electric field sets in when the moments are separated by at least 9 lattice spacings and persists to about 20 lattice spacings.  This window of tunability may be augmented by carefully adjusting $\mu$.  Regardless of the sign of the bias, once the band gap exceeds the chemical potential, the $\mu=0.05t$ curve begins to merge with the $\mu=0$ curve, as expected at low temperature.

This ability to dynamically tune the strength of the RKKY interaction, as well as
its sign from being antiferromagnetic to being ferromagnetic, for two fixed moments using just an electric field (rather than
doping) is perhaps the most interesting feature of this system.

\section{Discussion}
In this paper, we have discussed the physics of local moments on bilayer graphene. We have, in our discussion, ignored the effects of electron-electron interactions among the BLG electrons.  These are known to be important for quadratic band touching \cite{Sun:2009kx, Vafek:2010qy, Zhang:2010uq, Nandkishore:2010lr, Nandkishore:2010fj}, leading to symmetry breaking and many-body gaps for zero doping and zero electric field.  However, so long as there is nonzero doping or the presence of an electric field that gaps out the low-energy BLG states, the perturbative effects of electron-electron interactions are benign and will not lead to qualitatively new many-body effects.  Using a clean substrate may also be a viable route to mitigating the effects of rippling and disorder.  Moreover, the substrate may partially screen the BLG electron-electron interactions and reduce the many-body gap recently observed in suspended BLG \cite{Weitz05112010}.  The competition between impurity physics and many-body interactions in BLG deserves a careful separate investigation. 
The ability to turn on/off local moments, and the ability to tune the sign and magnitude of the RKKY coupling between local moments using electric fields perpendicular to the bilayer, which we have studied, constitutes physics beyond what has been discussed for monolayer graphene.  The experimental realization of such tunable local moments in BLG is a compelling prospect.   It would be interesting to study such a system using scanning tunnelling spectroscopy and to probe the quantum dynamics of interacting local moments in experiments. We expect that thermal fluctuations will only slightly alter the phase boundaries as long as the temperature is below the Hubbard gap.  Quantum fluctuations are expected to lead to Kondo screening or valence fluctuations in points of the phase diagram - this is an interesting direction for future research.

\ack
This research was supported by NSERC of Canada, an Ontario Early Researcher Award, and the DST (Government of India).
AP and MK acknowledge the hospitality of the Indian Institute of Science and the International Center for Theoretical Sciences while this manuscript was in preparation.

\bibliography{master}

\end{document}